\documentstyle[prd,aps,preprint]{revtex}
\tightenlines
\input epsf.tex
\def\DESepsf(#1 width #2){\epsfxsize=#2 \epsfbox{#1}}
\begin{document}

\draft
%@twocolumnfalse\endcsname
\preprint{\hbox{CTP-TAMU-34-99}}
\title{CP Violating Phases, Nonuniversal Soft Breaking And D-brane Models} 

\author{ E. Accomando, R. Arnowitt and B. Dutta }

\address{ Center For Theoretical Physics, Department of Physics, Texas A$\&$M
University, College Station TX 77843-4242}
\date{September, 1999}
\maketitle
\begin{abstract}The question of CP violating phases and electric dipole 
moments (EDMs)for the electron ($d_e$) and
the neutron ($d_n$) for supergravity models with
nonuniversal soft breaking is considered for models with a light 
($\stackrel{<}{\sim}$1 TeV) mass spectrum and R-parity invariance. As with
models with universal soft breaking (mSUGRA) one finds a serious fine tuning
problem generally arises for $\theta_{0B}$ (the phase of the B soft breaking
parameter at the GUT scale), if the experimental EDM constraints are obeyed and
radiative breaking of $SU(2)_L\times U(1)_Y$ occurs. A D-brane model where
$SU(3)_C\times U(1)_Y$ is associated with one set of 5-branes and $SU(2)_L$ with
another intersecting set of 5-branes is examined, and the cancellation phenomena is
studied over the parameter space of the model. Large values of $\theta_B$
(the phase of B at the electroweak scale) can be accommodated, though again
$\theta_{0B}$ must be fine tuned. Using the conventional prescription for
calculating $d_n$, one finds the region in parameter space where the
experimental EDM constraints on both $d_e$ and $d_n$ hold is significantly reduced, and
generally requires tan$\beta\stackrel{<}{\sim}$5 for most of the parameter space,
 though there are small allowed regions even for tan$\beta\stackrel{>}{\sim}$10.
 We find the Weinberg three gluon term generally makes significant
 contributions, and results are sensitive to the values of quark masses.
 \end{abstract}

\section{Introduction}
It has been realized for sometime that supersymmetric (SUSY) models allow for an
array of CP violating phases not found in the Standard Model(SM), and that these
phases will in general give rise to electric dipole moments (EDMs) of the
electron and the neutron which might violate the experimental bounds
\cite{all}.
The current 90 $\%$ C.L. bounds for $d_n$ and 95 $\%$ C.L. bounds for $d_e$
are quite stringent \cite{de} :
\begin{eqnarray}(d_n)_{exp}&<&6.3\times 10^{-26}ecm;\,\,\,\,(d_e)_{exp}<4.3\times
10^{-27}ecm\label{dnde}
\end{eqnarray}
 While these bounds can
always be satisfied by assuming sufficiently small phases (i.e O(10$^{-2}$))
and/or a heavy SUSY mass spectrum (i.e.$\stackrel{>}{\sim}$1 TeV) , recently it has been pointed out
that cancellations may occur allowing for ``naturally" large phases (i.e. O(10$^{-1}$))
and a light mass spectrum \cite{nath1} and this has led to considerable analysis both within
the MSSM framework \cite{nath1,kane1,pokorski} and the gravity mediated supergravity (SUGRA) GUT
framework \cite{nath2,nath3,falk1,bk,falk2,bartl,aad}. In the latter type models the theory is specified by
assigning the SUSY parameters at the GUT scale, and using the renormalization
group equations (RGEs), one determines the physical predictions at the electroweak
scale ($M_{\rm EW}$) (which we take here to be the t-quark mass,
$m_t$). Thus in SUGRA models, ``naturalness" is to be determined in terms of the
GUT parameters.

In a previous paper \cite{aad}, we examined the minimal model, mSUGRA, which
depends on the four universal soft breaking parameters at $M_G$ [$m_0$(scalar
mass), $m_{1/2}$ (gaugino mass), $A_0$ (cubic term mass) and $B_0$ (quadratic
term mass)] and the Higgs mixing parameter $\mu_0$. Since $m_0$ is real and we
can choose phases such that $m_{1/2}$ is real, one has only three phases at the
GUT scale in mSUGRA:
\begin{eqnarray} A_0&=&|A_0|e^{i \alpha_{0A}};\, B_0=|B_0|e^{i
\theta_{0B}};\,\mu_0=|\mu_0|e^{i
\theta_{0\mu}}.
\label{amb}
\end{eqnarray}
In Ref.\cite{aad}, it was shown that for the t-quark cubic soft breaking parameter
at $M_{\rm EW}$, $A_t=|A_t|e^{i\alpha_t}$, the nearness of the t-quark Landau
pole automatically suppresses $\alpha_t$ (the phase of $A_t$ at $M_{\rm EW}$),
 and one can satisfy the EDM bounds with
a light SUSY spectrum for large $\alpha_{0A}$, even $\alpha_{0A}=\pi/2$.
However, the situation is more difficult for $\theta_{0B}$. The experimental
requirements of Eq.(\ref{dnde}) combined with radiative breaking of
$SU(2)\times U(1)$ at $M_{\rm EW}$ imply that $\theta_{0B}$ is large i.e. O(1)
(unless $\alpha_{0A}$ is small and then all phases are small) and more serious, must
be tightly fine tuned unless tan$\beta$ is small (tan$\beta\stackrel{<}{\sim}$3). For example,
fixing $|A_0|$, $m_0$ and $m_{1/2}$ to be light and $\alpha_{0A}$ large, one
characteristically would find that $\theta_{0B}$ needs to be specified to 1 part
in 10$^4$ for tan$\beta$=10. Without this fine tuning, the GUT theory would not
achieve electroweak symmetry breaking at $M_{\rm EW}$ and/or satisfaction of
Eq.(\ref{dnde}).

Nonminimal models were also examined in \cite{aad} with results similar to the
above holding. In this paper we examine the nonminimal models in more detail.
We then discuss an interesting D-brane model \cite{kane2}, where the Standard Model
gauge group is associated with two 5-branes. This model results in nonuniversal
gaugino and scalar masses and is able to allow larger values of $\theta_B$ at
$M_{\rm EW}$. However, the same fine tuning problem at $M_G$ for $\theta_{0B}$
results in this model as well.

Our paper is organized as follows: Sec.2 reviews the basic formulae and notation
of the SUGRA GUT models for calculating the EDMs. Sec.3 examines a general class
of non universalities. Sec.4. considers the model of \cite{kane2} and conclusions are
given in Sec 5.

\section{EDMs in SUGRA Models} We consider here supersymmetry GUT models
possessing R-parity invariance where SUSY is broken in a hidden sector at a
scale above $M_G\cong 2\times 10^{16}$ GeV. This breaking is then transmitted by
gravity to the physical sector. The GUT group is assumed to be broken to the
Standard Model (SM) $SU(3)_C\times SU(2)_L\times U(1)_Y$ at $M_G$, but is
otherwise unspecified. The gauge kinetic function, $f_{\alpha\beta}$, and Kahler
potential, K, can then
give rise to nonuniversal gaugino masses at $M_G$ which we  parametrize by
\begin{eqnarray} m_{1/2i}&=&|m_{1/2i}|e^{i\phi_{0i}};\, i=1,2, 3.
\label{mphi}
\end{eqnarray}
 and we chose the phase convention where $\phi_{02}=0$. We also allow
nonuniversal Higgs and third generation masses at $M_G$ which can arise
from the Kahler potential:
\begin{eqnarray}
m_{H_1}^2&=&m_0^2(1+\delta_1);\,\,\,\,m_{H_2}^2=m_0^2(1+\delta_2)\\\nonumber
m_{q_L}^2&=&m_0^2(1+\delta_{3});\,m_{u_R}^2=m_0^2(1+\delta_4);
\,m_{e_R}^2=m_0^2(1+\delta_5);\\\nonumber
m_{d_R}^2&=&m_0^2(1+\delta_{6});\,m_{l_L}^2=m_0^2(1+\delta_7);
\label{scanonuni1}\end{eqnarray}
where $q_L\equiv(\tilde t_L,\,\tilde b_L)$, $u_R\equiv\tilde t_R$,
$e_R\equiv\tilde\tau_R$, etc., $m_0$ is the universal mass of the first two
generations and $\delta_i$ are the deviations from this for the Higgs bosons
and the third generation. In addition, there may be nonuniversal cubic soft
breaking parameters at $M_G$:
\begin{eqnarray} 
A_{0t}&=&|A_{0t}|e^{i\alpha_{0t}};\,A_{0b}=|A_{0b}|e^{i\alpha_{0b}};\,
A_{0\tau}=|A_{0\tau}|e^{i\alpha_{0\tau}}.
\end{eqnarray}

The electric dipole moment $d_f$ for fermion f is defined by the effective
Lagrangian:
\begin{eqnarray}  L_f&=&-{i\over 2}d_f\bar f\sigma_{\mu\nu}\gamma^5 f F^{\mu\nu}
\label{lag}
\end{eqnarray}
Our analysis follows that of \cite{nath1}. Thus the basic diagrams leading to the EDMs
are given in Fig.1. In addition there are gluonic operators
\begin{eqnarray}  L^G&=&-{1\over
3}d^Gf_{abc}G_{a\mu\alpha}G_{b\nu}^{\alpha}\tilde{G}_c^{\mu\nu}
\label{lg}
\end{eqnarray} and
\begin{eqnarray}  L^C&=&-{i\over 2}d^C\bar
q\sigma_{\mu\nu}\gamma^5T^aqG_a^{\mu\nu}
\label{lc}
\end{eqnarray} 
contributing to $d_n$ arising from the one loop diagram of Fig.1 (when the
outgoing vector boson is a gluon), the two loop Weinberg type diagram\cite{dai} 
and two loop Barr-Zee type
diagram\cite{ckp}. (In Eq.(\ref{lg}), $\tilde G^{\mu\nu}_{c}={1\over
2}\epsilon^{\mu\nu\alpha\beta}G_{c\alpha\beta}$, $\epsilon^{0123}=+1$, $T^a={1\over2}\lambda_a$, 
where  $\lambda_a$ are the
SU(3) Gellman matrices and $f_{abc}$ are the SU(3) structure constants).
We use naive dimensional analysis\cite{mon} to
relate these to the electric dipole moments and the QCD factors $\eta^{\rm
ED}$, $\eta^{G}$, $\eta^{C}$ to evolve these results to 1 GeV\cite{aln}. The quark
dipole moments are related to $d_n$ using the nonrelativistic quark model to
relate the u and d quark moments to $d_n$ i.e.
\begin{eqnarray}  d_n&=&{1\over 3}(4 d_d-d_u)
\label{dn}
\end{eqnarray}
and we assume the s-quark mass is 150 MeV. Thus QCD effects produce considerable
uncertainty in $d_n$ (perhaps a factor of 2-3).

Our matter phase conventions are chosen so that the chargino($\chi^{\pm}$),
neutralino($\chi^0$) and squark and slepton mass matrices take the
following form:
\begin{eqnarray} M_{\chi^{\pm}}&=&\left(\matrix{
 \tilde m_2                & \sqrt 2 M_W sin\beta  \cr
  \sqrt 2 M_W cos\beta            &-|\mu|e^{i\theta} }\right)
  \label{char}
\end{eqnarray}

\begin{eqnarray} M_{\chi^0}&=&\left(\matrix{
  |\tilde m_1| e^{i\phi_1}  &0            &a   &b\cr
  0          &\tilde m_2   &c   &d  \cr
  a          &c            &0   &|\mu|e^{i\theta}\cr
  b          &d            &|\mu|e^{i\theta}   &0  \cr}\right)
  \label{neut}
\end{eqnarray}
and
\begin{eqnarray} M_{\tilde q}^2&=&\left(\matrix{
 m^2_{q_L}               & e^{-i\alpha_q}m_q(|A_q|+|\mu| R_q
e^{i(\theta+\alpha_q)}) 
\cr
  e^{i\alpha_q}m_q(|A_q|+|\mu| R_q e^{-i(\theta+\alpha_q)})            
&m^2_{q_R} }\right).
\label{sqrk}
\end{eqnarray}
 In the above $a=-M_Z sin\theta_W cos\beta$, $b=M_Z sin\theta_W sin\beta$,
$c=-cot\theta_W a$, $d=-cot\theta_W b$,
$tan\beta=v_2/v_1$ ($v_{1,2}=\mid <H_{1,2}>\mid$), $R_q=cot\beta(tan\beta)$ for
u(d) quarks. All parameters are evaluated at the electroweak scale using the
RGEs, e.g. for quark q one has $A_q=|A_q|e^{i\alpha_q}$. (Similar formulae hold
for the slepton mass matrices.)

Electroweak symmetry breaking gives rise
to Higgs VEVs which we parametrize by
\begin{eqnarray}<H_{1,2}>&=&v_{1,2}e^{i\epsilon_{1,2}}\label{vev}.
\end{eqnarray}
These enter in the phase $\theta$ appearing in
Eqs.(\ref{char},\ref{neut},\ref{sqrk})
\begin{eqnarray}
\theta&\equiv&\epsilon_1+\epsilon_2+\theta_{\mu}\label{theta}
\end{eqnarray}
 The Higgs VEVs are calculated by minimizing the Higgs effective
 potential\cite{demir}:
\begin{eqnarray}
V_{eff}&=&m_{1}^2v_1^2+m_{2}^2v_2^2+2|B\mu|cos(\theta+\theta_B)v_1v_2+
{g^2_2\over 8}(v_1^2+v_2^2)^2+{{g^{\prime}}^2\over
8}(v_2^2-v_1^2)^2+V_1\label{veff}
\end{eqnarray} where $m^2_i=|\mu|^2+m^2_{H_i}$ and $m_{H_{1,2}}^2$ are the
Higgs running masses at $M_{\rm EW}$. $V_1$ is the one loop contribution.
\begin{eqnarray}V_1={1\over {64 \pi^2}}\sum_a C_a(-1)^{2 j_a} (2 j_a+1) m_a^4
(ln{m_a^2\over Q^2}-{3\over 2})\label{v1}
\end{eqnarray}  where $m_a$ is the mass of the a
particle of spin $j_a$, $Q$ is the electroweak scale (which we take to be
$m_t$) and $C_a$ is the number of color degrees of freedom. In the following we  include the full third generation states, $t$, $b$
and $\tau$ in
$V_1$ which allows us to treat the large tan$\beta$ regime. From Eq.(\ref{sqrk})
this implies that $V_1$ depends only on $\theta+\alpha_q$, $\theta+\alpha_l$
(though the rotation matrices which diagonalize $M_{\tilde q}^2$, $M_{\tilde
l}^2$ will depend on $\theta$, $\alpha_q$ and $\alpha_l$ separately). Minimizing $V_{eff}$
with respect to $\epsilon_1,\,
\epsilon_2$. then determines $\theta$:
\begin{eqnarray}\theta=\pi-\theta_B+
f_1(\pi-\theta_B+\alpha_q,\pi-\theta_B+\alpha_l )\label{thetaf}
\end{eqnarray}  where $f_1$ is the one loop correction. In general, $f_1$ is
small, but can become significant for large tan$\beta$, as discussed in
\cite{aad}.

Minimizing $V_{eff}$ with respect to $v_1$ and $v_2$ yields two equations which
can be arranged in the usual fashion to determine $|\mu|^2$ and $|B|$ at $M_{\rm
EW}$:
\begin{eqnarray}|\mu|^2&=&{{\mu^2_1-tan^2\beta\mu^2_2}\over{tan^2\beta-1}}-{1\over
2}M_Z^2\label{mumag}
\end{eqnarray} 
\begin{eqnarray}|B|&=&{1\over 2}sin{2\beta}{m_3^2\over|\mu|}\label{bmag}
\end{eqnarray} where
$\mu_i^2=m_{H_i}^2+\Sigma_i$, $m_3^2=2|\mu|^2+\mu^2_1+\mu^2_2$ and
$\Sigma_i=\partial V_1/{\partial v_i^2}$. Note that $|\mu|$ and $|B|$ depend
implicitly on the CP violating phases since the RGE that determines $m_{H_i}^2$
couple to the $A$ and $\tilde m_i$ equations, and $\Sigma_i$ depend on the
phases.

\section{NonMinimal Models} The renormalization group equations that relate
$M_{EW}$ to $M_G$ are in general complicated differential equations requiring
numerical solution, and all results given here are consequences of accurate
numerical integration. Approximate analytic solutions can however be found for
low and intermediate tan$\beta$ (neglecting b and
$\tau$ Yukawa couplings) and in the SO(10) limit of very large tan$\beta$
(neglecting the $\tau$ Yukawa coupling). These analytic solutions give some
insight into the nature of the more general numerical solutions.

For low and intermediate tan$\beta$, the $A_t$ and Yukawa RGEs read 
\begin{eqnarray}  -{{dA_t}\over {dt}}&=&6 Y_tA_t+{1\over {4
\pi}}(\sum^3_{i=1}a_i\alpha_i\tilde m_i)\\\nonumber -{{dY_t}\over {dt}}&=&6
Y_t-{1\over {4
\pi}}(\sum^3_{i=1}a_i\alpha_i)Y_t
\label{byt}
\end{eqnarray}
where $Y_t=h_t^2/{16\pi^2}$, $h_t$ is the t-quark Yukawa coupling constant and
$a_i$=(13/15,3,16/3). We follow the sign conventions of Ref.\cite{bop},
and $t=2ln(M_G/Q)$. The solutions of Eqs.(20) can be written as 
\begin{eqnarray}  A_t(t)&=&D_0A_{0t}-\tilde{H_2}+{{1-D_0}\over F} \tilde H_3
\label{at}
\end{eqnarray}
where 
\begin{eqnarray} 
\tilde{H_2}&=&{\alpha_G\over{4\pi}}t
\sum{{a_i|m_{1/2i}|e^{i\phi_i}}\over{1+\beta_it}}\equiv\sum_iH_{2i}|m_{1/2
i}|e^{i\phi_i}
\label{h2}
\end{eqnarray}

 and

\begin{eqnarray} 
\tilde{H_3}&=&\int^t_0{dt}'
E(t')\tilde{H_2}\equiv\sum H_{3i}|m_{1/2i}|e^{i\phi_i}
\label{h3}.\end{eqnarray} Here $D_0=1-6(F(t)/E(t))Y(t)$
vanishes at the t-quark Landau pole and hence is generally
small i.e. ($D_0\stackrel{<}{\sim}0.2$ for $m_t=175$ GeV). The functions F and E depend on the
SM gauge beta functions and are given in \cite{iba}. (E=12.3, F=254 for $t=2 ln
(M_G/m_t)$.) We note the identity\cite{iba}
\begin{eqnarray} 
{1\over F}\sum H_{3i}&=&t{E\over F}-1\cong 2.1
\label{h3f}.\end{eqnarray}
and so if we write Eq.(\ref{at}) as 
\begin{eqnarray} 
A_t(t)&=&D_0A_{0t}+\sum\Phi_i|m_{1/2i}|e^{i\phi_i}
\label{att}.\end{eqnarray}
the $\Phi_i$ are real and $O(1)$. (In the SO(10) large tan$\beta$ limit, one
obtains an identical result with the factor 6 replaced 7 in $D_0$. Thus
Eq.(\ref{att}) gives a valid qualitative picture over a wide range in
tan$\beta$.)

Nonuniversal gaugino masses affect the EDMs in two ways. First, taking the
imaginary part of Eq.(\ref{att}) one has ($\phi_2$=0 in our phase convention):

\begin{eqnarray} 
|A_t(t)|sin{\alpha_t}&=&|A_{0t}|D_0{\rm
sin}\alpha_{0t}+\sum_{i=1,3}\Phi_i|m_{1/2i}|{\rm sin}{\phi_i}
\label{astt}.\end{eqnarray}
As in the universal case, the smallness of $D_0$ suppresses the effects of any
large $\alpha_{0t}$ on the electroweak scale phase $\alpha_t$.
However large gaugino phases $\phi_i$ will generally make $\alpha_t$ large.
Second, Eqs.(\ref{neut}) and (\ref{char}) show that the phase $\phi_1$ enters into
the neutralino mass matrix though the chargino mass matrix remains unchanged
 ($\phi_2=0$). Thus the $\phi_1$ phase will affect
any cancellation occurring between the neutralino and chargino contributions to
the EDMs.

Some of the above efffects are illustrated in Figs. 2 and 3, where we plot K vs.
the phase $\theta_B$ at the electroweak scale for $d_e$. Here K is defined by 
\begin{eqnarray}  K=log_{10}\mid{d_f\over {(d_f)_{exp}}}\mid
\label{k}\end{eqnarray}
Thus K=0 corresponds to the case where the theory saturates the current
experimental EDM bound, while K=-1, would be the situation if the experimental
bounds were reduced by a factor of 10. Fig.2 considers universal scalar masses
and universal $A_0$ with $\alpha_{0A}=\pi/2$ at the GUT scale, and
$\phi_1=\phi_3=-1.1\pi$, -1.3$\pi$, -1.5$\pi$ for tan$\beta$=3. 
We see that as $|\phi_1|$
is increased from $|\phi_1|$=1.1$\pi$ to 1.3 $\pi$, the allowed values of 
$\theta_B$
increases significantly since the $\phi_1$ phase in Eq. (11) aids the cancellation
between the neutralino and the chargino contributions. However, 
increasing $|\phi_1|$
further to $|\phi_1|$=1.5$\pi$ over compensates causing the allowed values of 
$\theta_B$
to decrease. Fig. 3 for tan$\beta$=10 shows a similar effect. The experimentally allowed
parameters require $K\leq 0$. The allowed range $\Delta\theta_B$ of $\theta_B$ decreases
with tan$\beta$. It is very small for tan$\beta$=10 and is quite small even for tan$\beta$=3.
\section{D-Brane Models}
Recent advances in string theory leading to possible D=4, N=1 supersymmetric
vacua after compactification has restimulated interest in phenomenological
string motivated model building. A number of approaches exists including models
based on Type IIB orientifolds, Horava-Witten M theory compactification on
$CY\times S^1/Z_2$ and perturbative heterotic string vacua. The existence of
open string sectors in Type IIB strings implies the presence of $Dp$-branes,
manifolds of p+1 dimensions in the full D=10 space of which 6 dimensions are
compactified e.g. on a six torus $T^6$. (For a survey of properties of Type IIB
orientifold models see \cite{iba2}). One can build models containing 9 branes (the
full 10 dimensional space) plus $5_i$-branes, i=1, 2, 3 (each containg two of
the compact dimensions) or 3 branes plus 7$_i$ branes, i=1, 2, 3 (each having
two compactified dimensions orthogonal to the brane). Associated with a set of
n coincident branes is a gauge group U(n). Thus there are large number of ways
one might embed the Standard Model gauge group in Type IIB models.

We consider here an interesting model recently proposed \cite{kane2} based on
9-branes and 5-branes. In this model, $SU(3)_C\times U(1)_Y$ is associated
with one set of 5-branes, i.e. $5_1$, and SU(2)$_L$ is associated with a
second intersecting set 5$_2$. Strings starting on 5$_2$ and ending on 5$_1$
have massless modes carrying the joint quantum numbers of the two branes (we
assume these are the SM quark and lepton doublets, Higgs doublets) while
strings beginning and ending on 5$_1$ have massless modes carrying
$SU(3)_C\times U(1)_Y$ quantum numbers (right quark and right lepton states). A
number of general properties of such models have been worked out \cite{iba2}. Thus
to accommodate the phenomenological requirement of gauge coupling constant
unification at $M_G\cong 2\times 10^{16}$ GeV, one may assume, that $M_c$,
the compactification scale of the Kaluza-Klein modes obeys $M_c=M_G$. Above
$M_c$, the gauge interactions on the 5-branes see a D=6 dimensional space (with Kaluza
Klein modes)
while above $M_c$ gravity sees the full D=10 space. Gravity and gauge
unification then is to take place at the string scale $M_{\rm
str}=1/{\sqrt{\alpha'}}$ given by $M_{\rm str}=(\alpha_G M_c
M_{Planck}/\sqrt{2})^{1/2}\cong8\times 10^{16}$ GeV (for $\alpha_G\cong1/24$).

The gauge kinetic functions for 9 branes and 5$_i$-branes are given by
\cite{iba2,iba1}
$f_9=S$ and $f_{5_i}=T_i$ where S is the dilaton and $T_i$ are moduli. The
origin of the spontaneous breaking of N=1 supersymmetry and of compactification is not yet understood
within this framework. Further, CP violation must also occur as a spontaneous
breaking.  One assumes these effects can be phenomenologically accounted for
by F-components growing VEVs parametrized as \cite{iba2,iba3,iba4}
\begin{eqnarray} F^S&=&{2\sqrt{3}}<{\rm Re} S>{\rm
sin}\theta_b e^{i\alpha_s}m_{3/2}\\\nonumber 
F^{T_i}&=&{2\sqrt{3}}<{\rm Re} T_i>{\rm
cos}\theta_b \Theta_i e^{i\alpha_i}m_{3/2}
\label{fsft}
\end{eqnarray}
where $\theta_b$, $\Theta_i$ are Goldstino angles
($\Theta_1^2+\Theta_2^2+\Theta_3^2=1$). CP violation is thus incorporated in
the phases $\alpha_s$, $\alpha_i$. In the following we will assume, for
simplicity, that $\Theta_3=0$ (i.e. that the 5$_3$-brane does not affect the
physical sector). We also assume isotropic compactification ($<Re T_i>$) are
equal) to guarantee grand unification at $M_G$, and $<Im T_i>$=0 so that the
spontaneous breaking does not grow a $\theta$-QCD type term.

For models of this type, T-duality determines the Kahler potential
\cite{iba2,iba3,iba4} and,
combined with Eq.(28), generates the soft breaking terms. One finds at
$M_G$\cite{iba2,iba3,iba4}:
\begin{eqnarray}  \tilde{m_1}&=&\sqrt{3}{\rm
cos}\theta_b\Theta_1e^{-i\alpha_1}m_{3/2}=\tilde{m_3}=-A_0\\
\tilde{m_2}&=&\sqrt{3}{\rm
cos}\theta_b\Theta_2e^{-i\alpha_2}m_{3/2}
\label{m1m2}
\end{eqnarray}
and
\begin{eqnarray}  m_{5_15_2}^2&=&(1-{3\over 2}{\rm sin}^2\theta_b) m^2_{3/2}\\
 m_{5_1}^2&=&(1-{3}{\rm sin}^2\theta_b) m^2_{3/2}
\label{m512}
\end{eqnarray}
Here $A_0$ is a universal cubic soft breaking mass, $m_{5_15_2}^2$ are the
soft breaking masses for $q_L$, $l_L$, $H_{1,2}$ and $m_{5_1}^2$ are for
$u_R$, $d_R$ and $e_R$.

We see that the brane models give rise to nonuniversalities that are
strikingly different from what one might expect in SUGRA GUT models. Thus it
would be difficult to construct a GUT group, which upon spontaneous breaking at
$M_G$ yields gaugino masses $\tilde m_1$=$\tilde m_3\neq \tilde m_2$, and
similarly the above pattern of sfermion and Higgs soft masses. Brane models
can achieve the above pattern since they have the freedom of associating different
parts of the SM gauge group with different branes. 

The above model does not determine the $B$ and $\mu$ parameters. We therefore
will phenomenologically parametrize these at $M_G$  by
\begin{eqnarray}  B_0=|B_0|e^{i
\theta_{0B}};\,\mu_0=|\mu_0|e^{i
\theta_{0\mu}}.
\label{mb}
\end{eqnarray}
with two additional CP violating phases $\theta_{0B}$ and $\theta_{0\mu}$. We
also set $\alpha_2=0$ in the following. 

We consider first the electron EDM. (We use the interactions of Ref.\cite{gh}
including the Erratum on the sign of Eq.(5.5) of Ref.\cite{gh}). Fig.4 plots K as a function of $\theta_B$
for tan$\beta$=2 (solid), 5 (dashed), 10 (dotted) with phases
$\phi_1=\phi_3=\pi+\alpha_{0A}=-1.25\pi$ and $m_{3/2}=150$ GeV, $\theta_b=0.2$,
$\Theta_1$=0.85. We see that the EDM bounds allow remarkably large values of
$\theta_B$ in this model even for large tan$\beta$, e.g. $\theta_B\cong0.4$ for
tan$\beta$=2 and $\theta_B\cong0.25$ for
tan$\beta$=10. (A second allowed region occurring approximately for
$\theta_B\rightarrow\pi+\theta_B$ also exists. However this corresponds to the
sign of $\mu$ that is mostly excluded by the $b\rightarrow s\gamma$ data.) Fig.5 shows a similar plot for somewhat smaller phases
$\phi_1=\phi_3=\pi+\alpha_{0A}=-1.1\pi$. Again relatively large phases can exist
at the electroweak scale.

As discussed in Ref.\cite{kane2}, the largeness of $\theta_B$ is due to an enhanced
cancellation between the neutralino and chargino contributions as a consequence
of the additional $\phi_1$ dependence in Eq.(11), allowing $\theta_B$ to be
O(1). However, in spite of this, the range in $\theta_B$ at the electroweak
scale, where the experimental bound $K\leq 0$ is satisfied, is quite small, e.g.
from Fig.4, $\Delta\theta_B\simeq$0.015 even for tan$\beta$=2. As discussed in
Ref\cite{aad}, this implies that the radiative breaking condition makes the
allowed range $\Delta\theta_{0B}$ at the GUT scale very small, particularly for
large tan$\beta$. This is illustrated in Figs. 6 and 7. In Fig.6, we have
plotted the central value of $\theta_{0B}$ which satisfies $K\le 0$ as a
function of tan$\beta$. Thus $\theta_{0B}$ is generally quite large. In Fig.7
we have plotted the allowed range of $\Delta\theta_{0B}$ satisfying the EDM
constraints. One sees that even for small tan$\beta$ the
allowed range $\Delta\theta_{0B}$ is very small. Thus as in the mSUGRA model of
Ref.\cite{aad}, one has a serious fine tuning problem at the GUT scale due to the
combined conditions of radiative breaking and the EDM bound:
$\theta_{0B}$ must be large but very accurately determined by the string model
if it is to agree with low energy phenomenology.

The neutron dipole moment is more complicated due to the additional
contributions arising from $L^C$ and $L^G$ of Eqs.(8) and (7). While there are
significant uncertainties in the calculation of $d_n$ it is of interest to see
if the experimental bounds on $d_n$ can be achieved in the same region of
parameter space as occur in $d_e$ above. The fact that the brane model requires
$\phi_3=\phi_1$ allows for the $L^C$ gluino contribution to aid in canceling
the chargino contribution. This generally aids in broadening the overlap
region of joint satisfaction of the $d_n$ and $d_e$ bounds of Eq.(\ref{dnde}).
 However, in
addition to this, there is a contribution from $L^G$ from the Weinberg type
diagram. While this term is enhanced due to the factor of $m_t$, it is a two
loop diagram and is suppressed by a factor of $\alpha_3^2$($g_3/{4\pi}$) and
in most models is usually a small contribution. However, for the D-brane model
where $\phi_3=\phi_1$, the presence of a large $\phi_3$ phase increases the
significance of this diagram, reducing the $d_n-d_e$ overlap region.
 This is
illustrated in Figs.8 where $\Theta_1$ is plotted as a function of $\theta_B$
for parameters tan$\beta$=2, $m_{3/2}=150$ GeV, $\theta_b$=0.2. (LEP189 bounds
of $m_{1/2}\stackrel{>}{\sim} 150$ GeV imply here that $\Theta_1\stackrel{<}{\sim}0.94$.) As one proceeds from 
$\phi_1=\phi_3=-1.25\pi$ of Fig.8a to $\phi_1=\phi_3=-1.95\pi$ of Fig.8d, one
goes from no overlap of the allowed $d_e$ and $d_n$ regions to a significant
overlap. However, the large 
$\theta_B$ phase  allowed separately by $d_e$ and $d_n$ (e.g. $\theta_B\sim
0.6$) in Fig.8a is sharply reduced in Fig.8d in the overlap region by a factor
of 10.  Further, the region of parameter space
where the experimental constraints for $d_e$ and $d_n$ can be
simultaneously satisfied generally decreases with increasing tan$\beta$. Fig.9
gives the allowed region for the parameters of Fig. 8b with
$\phi_3=\phi_1=-1.90\pi$, for tan$\beta$=2, 3 and 5. The allowed parameter
 space disappears for tan$\beta\stackrel{>}{\sim}5$. If, however, the overlap in allowed
parameter region between $d_e$ and $d_n$ occurs for smaller $\phi_1$ i.e.
$\phi_1=O(10^{-1})$, one can have larger values of tan$\beta$. This is illustrated in
Fig.10 for $\phi_1=\phi_3=-1.97\pi$ (i.e. 2$\pi+\phi_1$=0.03$\pi$) 
for tan$\beta$=10. The region of
overlap however now requires $\theta_B$ to be quite small i.e.
$\theta_B=O(10^{-2})$. Of course the fine tuning of $\theta_{0B}$ at the
GUT scale becomes quite extreme for larger tan$\beta$ \cite{25}. 

While the quark mass ratios are well determined, the values of $m_u$ and $m_d$
remain very uncertain due to the uncertainty in $m_s$ \cite{leut}. As pointed
out in Ref.\cite{aad}, this uncertainty contributes significantly to the
uncertainty in the calculation of $d_n$. This effect for the model of
Ref.\cite{kane2} is illustrated in Fig.11 for $\phi_1=\phi_3=-1.90\pi$, 
tan$\beta$=2.
Fig. 11a corresponds to the choice of light quarks ($m_s({\rm 1 GeV})\simeq$ 95 MeV) while
Fig. 11b to heavy quarks ($m_s({\rm 1 GeV})\simeq$ 225 MeV). For light quarks, the Weinberg
three gluon term makes a relatively larger contribution and aids more in the
cancellation needed to satisfy the EDM constraint. In general, though, the
Weinberg term can be several times the upper bound on $d_n$ of Eq.(1), and so
makes a significant contribution. In other figures of this paper, we have used a
central value of $m_s$ i.e $m_s({\rm 1 GeV})=150$ MeV corresponding to $m_u({\rm 1 GeV})\simeq 4.4$ MeV
and $m_d({\rm 1 GeV})\simeq 8$ MeV.

\section{Conclusions}In minimal SUGRA models with universal soft breaking, it
has previously been seen that the current EDM constraints can be satisfied
without fine tuning the CP violating phases at the electroweak scale. For this
case the EDMs are most sensitive to $\theta_B$, the phase of the B parameter, and
experiment can be satisfied with $\theta_B$=O(10$^{-1}$). It was seen however
that at the GUT scale, $\theta_{0B}$ was generally large (unless masses were
large or the other phases were small), and in order to satisfy both the EDM
constraints and radiative electroweak breaking, $\theta_{0B}$ had to be fine
tuned, the fine tuning becoming more serious as tan$\beta$ increased\cite{aad}. In
this paper we have examined nonuniversal models, and have found generally that
the same phenomenon exists. 

We have studied in some detail an interesting D-brane model involving CP
violating phases where the Standard Model gauge group is embedded on two sets of
5-branes, $SU(3)_C\times U(1)_Y$ on $5_1$ and $SU(2)_L$ on $5_2$ so that the
gaugino phases obey $\phi_3=\phi_1\neq\phi_2$\cite{kane2}. This is a symmetry
breaking pattern that is different from what one normally expects in GUT models. If
one examines $d_e$ and $d_n$ separately, one finds that this model can
accommodate remarkably large values of $\theta_B$ i.e. $\theta_B$ as large as
 0.7.
However, the same fine tuning problem arises at the GUT scale for
$\theta_{0B}$. Further, the region in parameter space where the experimental
bounds on both $d_e$ and $d_n$ are satisfied shrinks considerably. 
Thus the model can not actually realize
the very largest $\theta_B$ (though $\theta_B$ as large as $\simeq 0.4$ is still
possible). The Weinberg three gluon diagram typically is several times the
current experimental upper bound on $d_n$, and so makes a significant
contribution, particularly if the quark masses are light. (The Barr-Zee term is
generally small if the SUSY parameters are $\stackrel{<}{\sim}1$ TeV.) The allowed region in parameter space which simultaneously
satisfies the $d_n$ and $d_e$ constraints also shrinks as tan$\beta$ 
is increased,
the $d_e$ and the $d_n$ allowed regions narrowing. In general, if one assumes large
$\phi_i$ phases, one needs tan$\beta\stackrel{<}{\sim}$5 to get a significant
overlap between the allowed
$d_e$ and allowed $d_n$ regions in parameter space, though small overlap
regions exist even for tan$\beta=10$ and higher (though with
$\theta_B=O(10^{-2})$). In the search for the SUSY
Higgs, the Tevatron in RUN II/III will be able to explore almost the entire
region of tan$\beta\stackrel{<}{\sim}$50 (for SUSY parameters 
$\stackrel{<}{\sim}$1 TeV)\cite{lep} and it should be possible to experimentally verify
whether tan$\beta$ is in fact small.

As commented in Sec.2, the theoretical calculation of $d_n$ contains a number
of uncertainties due to QCD effects. We have used here the conventional
analysis. However, these uncertainties could affect the overlap between the
allowed $d_e$ and $d_n$ regions, and modify bounds on $\theta_B$ and tan$\beta$.
However, if QCD effects are not too large, we expect the general features described
above to survive.
\section{Acknowledgement}

This work was supported in part by National Science Foundation Grant No.
PHY-9722090. We should like to thank M. Brhlik for discussions of the results of
Ref.\cite{kane2}, and JianXin Lu for useful conversations.

\begin{figure}[htb]
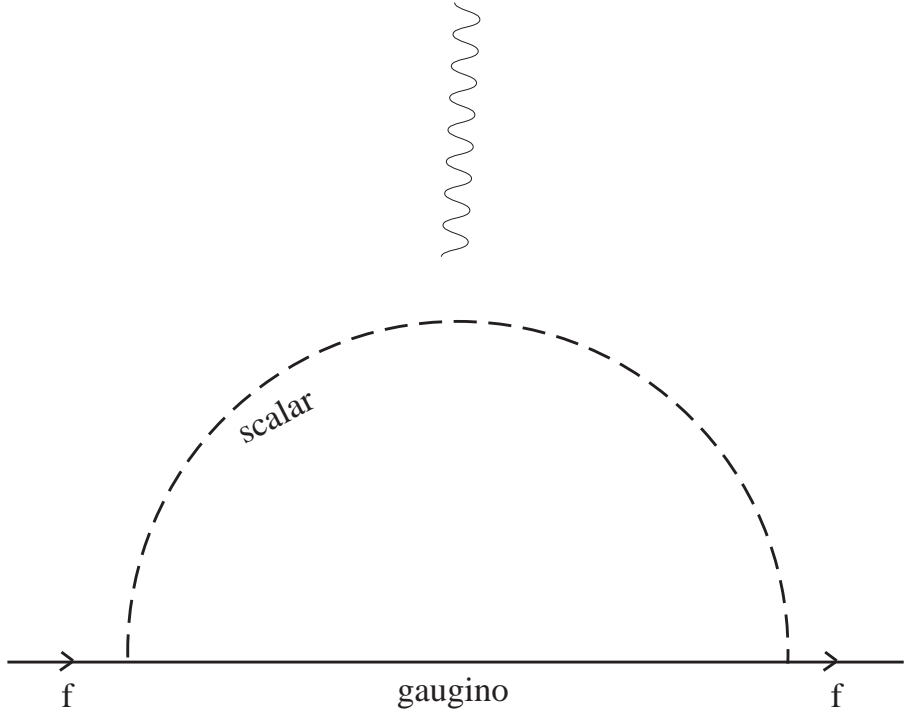

\centerline{ \DESepsf(edmfig1.epsf width 12 cm) }
\smallskip
\caption {\label{fig1} One loop diagram. The photon line can be attached
to any charged particle.}
\end{figure}
\begin{figure}[htb]
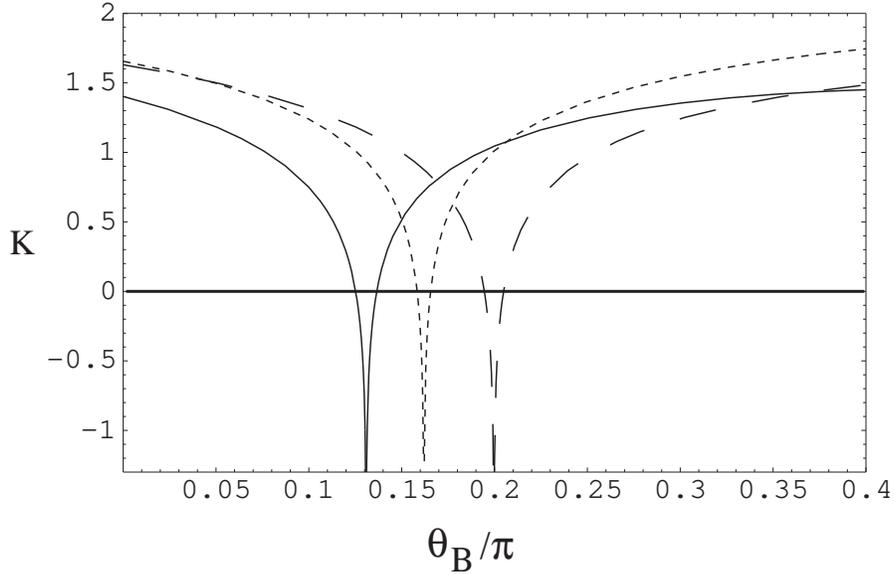

\centerline{ \DESepsf(aad2edmfig2.epsf width 12 cm) }
\smallskip
\caption {\label{fig2} K vs. $\theta_B$ for  $d_e$ for $m_0$=100 GeV, 
$m_{1/2}$=200 GeV, $|A_0|$=300 GeV, $\alpha_{0A}=\pi/2$,
 $\phi_1$=$\phi_3$=-1.1$\pi$(solid), -1.3$\pi$
(dashed), -1.5 $\pi$ (dotted) and tan$\beta$=3.}
\end{figure}
\begin{figure}[htb]
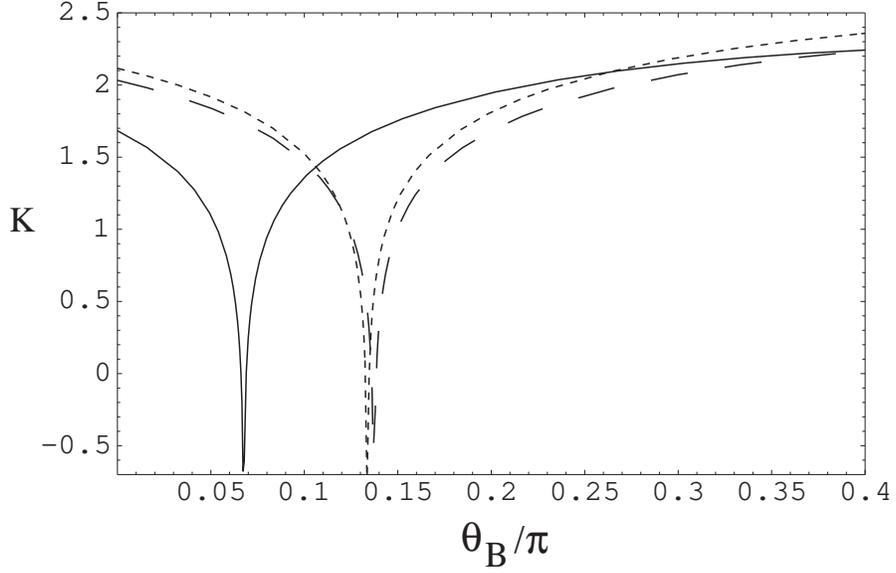

\centerline{ \DESepsf(aad2edmfig3.epsf width 12 cm) }
\smallskip
\caption {\label{fig3} same as Fig.2 for tan$\beta$=10.}
\end{figure}
\begin{figure}[htb]
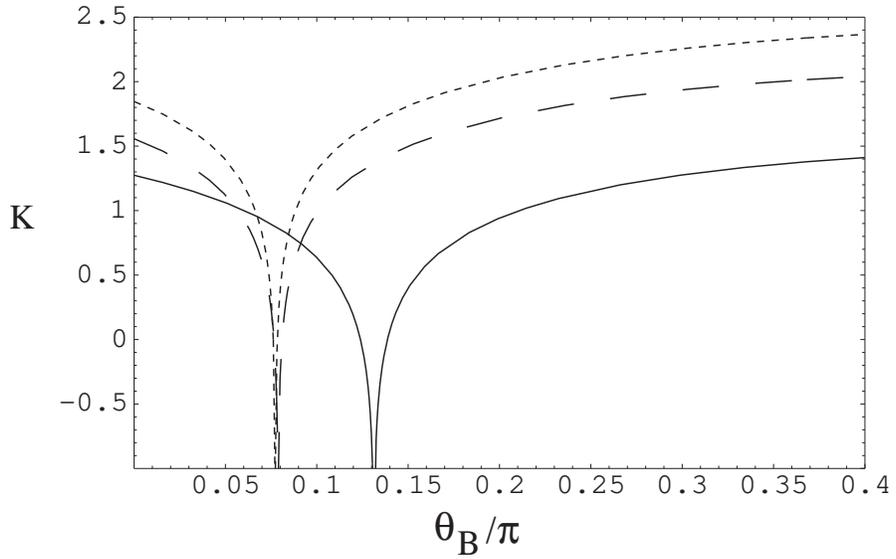

\centerline{ \DESepsf(aad2edmfig5.epsf width 12 cm) }
\smallskip
\caption {\label{fig4} K vs. $\theta_B$ for $d_e$ for 
$\phi_1$=$\phi_3$=$\pi+\alpha_{0A}$=-1.25$\pi$, 
$m_{3/2}$=150 GeV, $\theta_b$=0.2, $\Theta_1=0.85$, with tan$\beta$=2(solid), 5(dashed),
10(dotted).}
\end{figure}
\begin{figure}[htb]
\centerline{ \DESepsf(aad2edmfig4.epsf width 12 cm) }
\smallskip
\caption {\label{fig5} same as Fig.4 for 
$\phi_1$=$\phi_3$=$\pi+\alpha_{0A}$=-1.1$\pi$.}
\end{figure}
\begin{figure}[htb]
\centerline{ \DESepsf(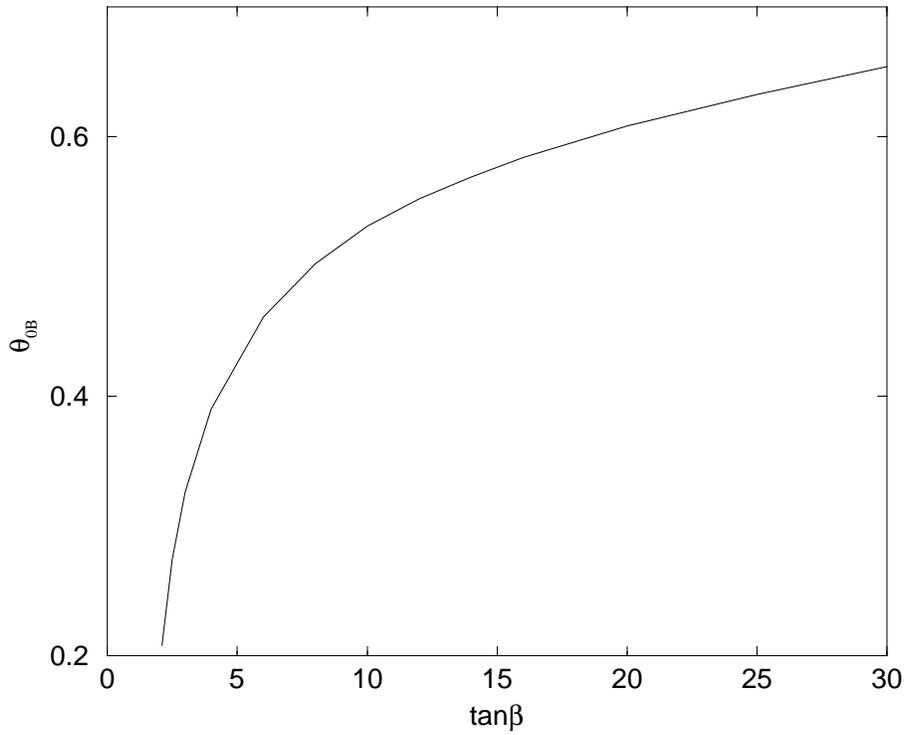 width 12 cm) }
\smallskip
\caption {\label{fig6} Central values of $\theta_{0B}$ for $d_e$ satisfying the EDM
constraint as a function of tan$\beta$. Input parameters are as in
Fig.4.}
\end{figure}\begin{figure}[htb]
\centerline{ \DESepsf(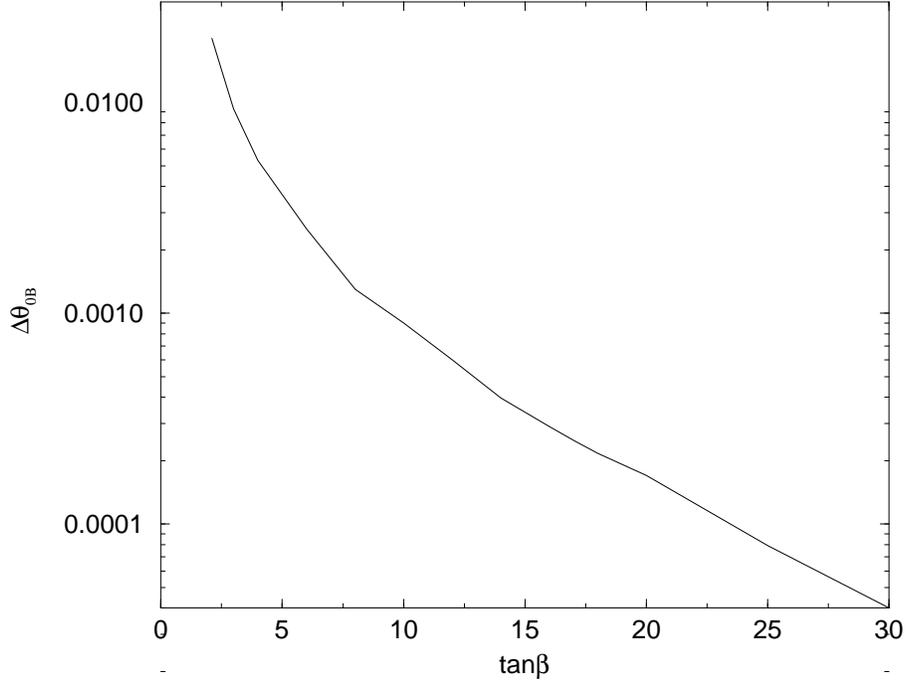 width 12 cm) }
\smallskip
\caption {\label{fig7} Values of $\Delta\theta_{0B}$ for $d_e$ satisfying the EDM
constraint as a function of tan$\beta$. Input parameters are as in
Fig.4.}
\end{figure}
\begin{figure}[htb]
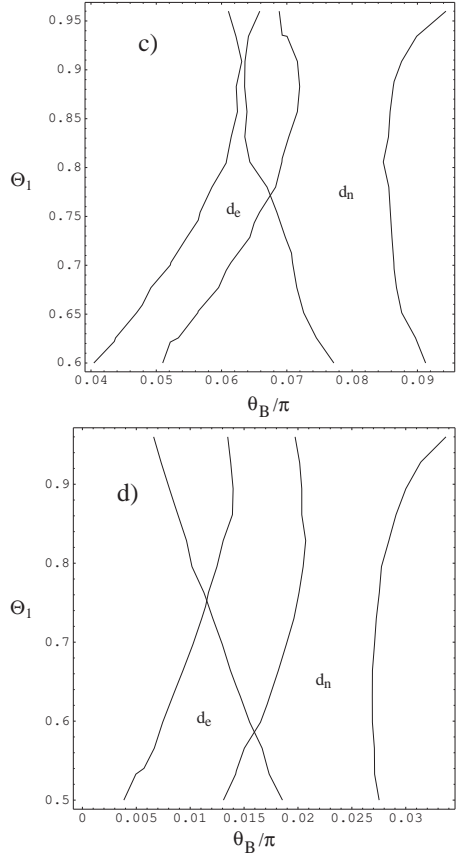

\centerline{ \DESepsf(aad2edmfig2125new.epsf width 6 cm) }
\centerline{ \DESepsf(aad2edmfig2160new.epsf width 6 cm) }
\centerline{ \DESepsf(aad2edmfig218new.epsf width 6 cm) }
\centerline{ \DESepsf(aad2edmfig2195new.epsf width 6 cm) }
\smallskip
\caption {\label{fig8} Allowed regions for $d_e$ and $d_n$
 for tan$\beta$=2, $\theta_b$=0.2 and $m_{3/2}$=150 GeV.
a) $\phi_1$=$\phi_3$=-1.25$\pi$, b) $\phi_1$=$\phi_3$=-1.60$\pi$, c)
$\phi_1$=$\phi_3$=-1.80$\pi$ and d)
$\phi_1$=$\phi_3$=-1.95$\pi$. }
\end{figure}
\begin{figure}[htb]
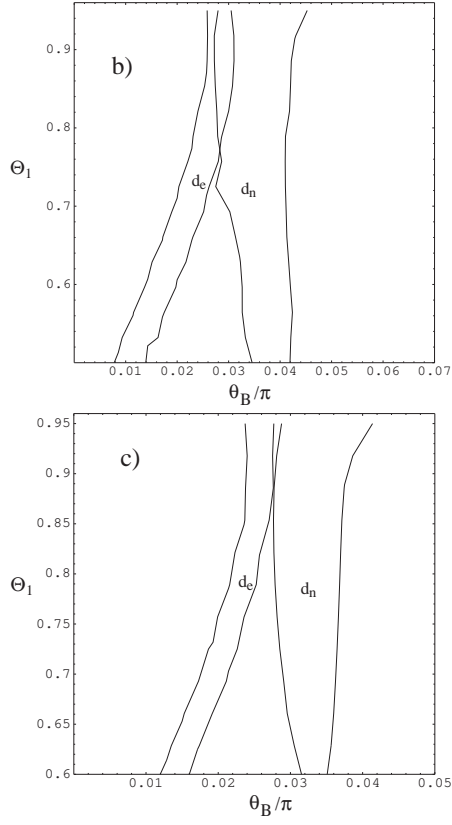

\centerline{ \DESepsf(aad2edmfig219new.epsf width 6 cm) }
\centerline{ \DESepsf(aad2edmfig319new.epsf width 6 cm) }
\centerline{ \DESepsf(aad2edmfig519new.epsf width 6 cm) }
\smallskip
\caption {\label{fi9}Allowed regions for $d_e$ and $d_n$
  for $\theta_b$=0.2, $m_{3/2}$=150 GeV and
 $\phi_1$=$\phi_3$=-1.90$\pi$, for a) tan$\beta$=2, b) tan$\beta$=3 and c)
tan$\beta$=5.}
\end{figure}
\begin{figure}[htb]
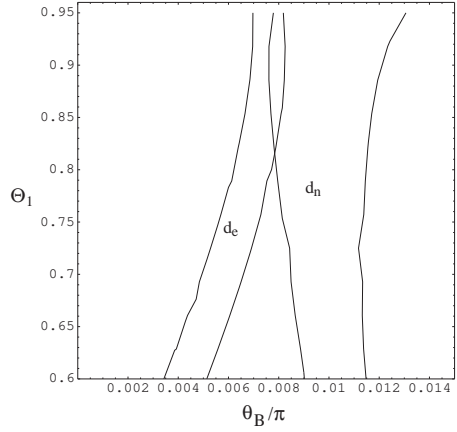

\centerline{ \DESepsf(aad2edmfig10197new.epsf width 6 cm) }
\smallskip
\caption {\label{fig10}Allowed regions for $d_e$ and $d_n$
  for $\theta_b$=0.2, $m_{3/2}$=150 GeV and
 $\phi_1$=$\phi_3$=-1.97$\pi$, for  tan$\beta$=10.}
\end{figure}
\begin{figure}[htb]
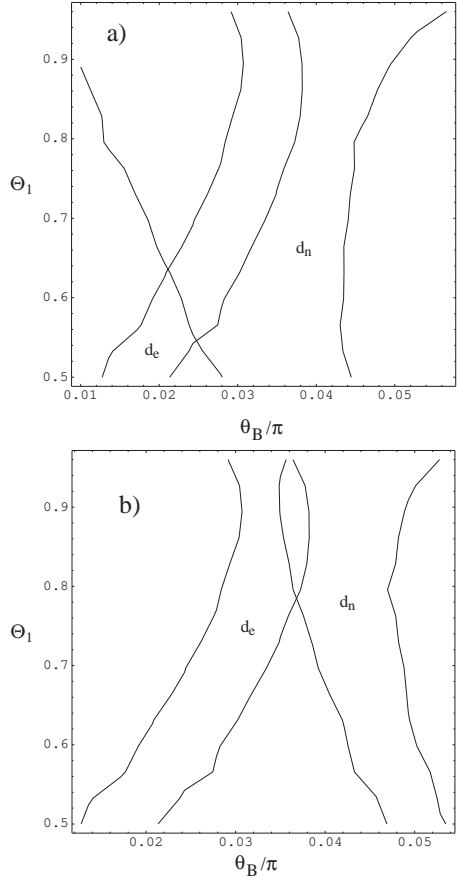

\centerline{ \DESepsf(aad2edmfig219udlo.epsf width 6 cm) }
\centerline{ \DESepsf(aad2edmfig219udhi.epsf width 6 cm) }
\smallskip
\caption {\label{fig11}Allowed regions for $d_e$ and $d_n$
  for $\theta_b$=0.2, $m_{3/2}$=150 GeV, tan$\beta=2$ and
 $\phi_1$=$\phi_3$=-1.90$\pi$ for a) $m_u$(1 GeV)=2.75 MeV, $m_d$(1 GeV)=5.0 MeV and 
 b) $m_u$(1 GeV)=6.65 MeV, $m_d$(1 GeV)=12 MeV.}
\end{figure}

\begin{thebibliography}{99}
\bibitem{25} The above results for tan$\beta$=2  are in general agreement
with the corrected calculations of the results of Ref.\cite{kane2}.
\end{thebibliography}
\end{document}